\documentclass[aps,pra,twocolumn,footnote]{revtex4}
\usepackage{amsfonts}
\usepackage{amssymb}
\usepackage{color}
\usepackage{amsmath}
\usepackage{graphicx}
\usepackage{subfigure}
\usepackage{txfonts}
\usepackage{bm}
\usepackage{ulem}
\usepackage[colorlinks,linkcolor=blue,citecolor=blue]{hyperref}
\usepackage{ulem}

\newcommand{\hz}[1]{{\color{blue}#1}}
\newcommand{\PT}{$\mathcal{PT}$}
\newcommand{\nh}{non-Hermitian }
\newcommand{\hatsigma}{\hat{\sigma}}
\newcommand{\identity}{\mathbb{I}}
\newcommand{\steadystate}{\rm ss}

\begin{document}

\title{Parameter estimation with the steady states of non-Hermitian spin chains }

\author{Huiqin Zhang and Jiasen Jin}
\email{jsjin@dlut.edu.cn}
\affiliation{School of Physics, Dalian University of Technology, Dalian 116024, China}

\begin{abstract}
We propose a scheme for parameter estimation with the steady states of non-Hermitian spin chains. The parameters to be estimated are encoded in the system via the external magnetic field that imposed on the first site of the chain. We obtain the analytical spectrum for the spin chain of two sites. We find that the quantum Fisher information (QFI) about the amplitude of the imposing field diverges while the QFI about the azimuthal angle reaches to the maximum at exceptional points. Moreover, the QFI is enhanced as the system size increasing and saturates for sufficiently long spin chain because only short-range correlations are induced by the imposing field.
\end{abstract}
\date{\today}
\maketitle

\section{Introduction}
In quantum mechanics, the Hermiticity of observables guarantees the outcome of measurements to be real. In particular\hz{,} the real eigenvalues of Hermitian Hamiltonian lead the time-evolution of the states of closed quantum systems to be unitary. However, in recent years the \nh systems have attracted increasing attention in the areas of quantum information \cite{xiao2019,zhan2020}, quantum optics \cite{jing2014,qi2020,wang2021}, quantum phase transitions \cite{lee2014a,lee2014b,kawabata2017,yao2018,zhangkl2021} and topological phases \cite{yao2018b,gong2018,kawabata2019}. A comprehensive panorama of the studies on \nh physics can be found in Ref. \cite{ashida2020}. On the one hand, it is known that the \nh Hamiltonians with parity-time ({\PT}) symmetry still possess real spectra \cite{bender1998}. The \PT-symmetric systems are of great interest in nonreciprocal light transport \cite{bi2011,feng2021,wang2013,peng2021}, mode switching \cite{xu2016}, and quantum sensing \cite{liu2016,yi2021}. On the other hand, the complex eigenvalues of \nh Hamiltonians break the unitary time-evolution of the state of quantum systems. Moreover, the combination of \nh Hamiltonian and the quantum jumps provide a reliable description for the dynamics of open quantum systems \cite{plenio1998}. One of the most notable processes is the spontaneous emission of atoms in which the time-evolution of a two-level atom is under continuous detection of the emitting photon. When the detector is not triggered the wave function evolves according to an effective \nh Hamiltonian while a quantum jump collapses the atom to the ground state once the detector counts \cite{carmichael1993}. The nonzero (always non-positive) imaginary parts of the eigenvalues of  \nh Hamiltonian always drive the quantum system to a steady state in long-time limit. The steady state is the eigenstate associated to the eigenvalue with the maximal imaginary part of the \nh Hamiltonian, namely the eigenstate decays slowest during the time-evolution according to the \nh Hamiltonian.

Analogous to the ground-state phase transitions in Hermitian quantum systems, the steady states of \nh quantum systems may also undergo phase transitions with symmetry breaking by tuning the controllable parameter of the Hamiltonian across the exceptional point (EP) \cite{lee2014a,lee2014b}. The EP indicates coalescence of both the eigenvalues and the eigenstates of the \nh Hamiltonian. As the system approaches to the EP (or critical point) of steady-state phase transition, the steady-state entanglement, correlations as well as the response function exhibit nonanalytical singular behaviors. The singularities of the \nh quantum system can be utilized for quantum parameter estimation \cite{giovannetti2004,giovannetti2011}.

Quantum parameter estimation is an important task in quantum information and quantum metrology. Generally, the value of a parameter $\eta$ in a quantum system can be estimated through the following steps: Firstly one prepares a quantum sensor which can be a carefully engineered quantum system with Hamiltonian $\hat{H}_0$. Secondly, the parameter $\eta$ is encoded into the quantum sensor by introducing a perturbing Hamiltonian $\hat{H}_1(\eta)$ to the system. Then the value of $\eta$ is inferred by measuring the appropriate observable of the systems with Hamiltonian $\hat{H}(\eta)=\hat{H}_0+\hat{H}_1(\eta)$. So far a large number of schemes for quantum parameter estimations have been proposed in quantum system with \PT-symmetric \cite{lau2018,mcdonald2020}, pseudo-Hermitian Hamiltonian \cite{chu2020}, and non-Markovian dissipations \cite{wu2021a,wu2021b,zhang2021}.  Schemes with time-dependent Hamiltonians have also been proposed \cite{pang2014,pang2017,chu2021b}. Assuming that the estimation is asymptotically locally unbiased, the estimation precision can be characterized by the distinguishability between the outcome of measurements involving two neighboring parameters which is constrained by the well-known Cram\'{e}r-Rao bound \cite{Cramer1946, Wootters1981},
\begin{equation}
\delta\eta\ge\frac{1}{\sqrt{\nu F_\eta}},
\label{QCRB}
\end{equation}
where $\delta\eta$ is the root-mean-square deviation, $\nu$ is the rounds of measurements and $F_\eta$ is the Fisher information. The asymptotically locally unbiased assumption for which Eq. (\ref{QCRB}) holds means that the estimation of $\eta$ tends to the correct value in the limit of $\nu\rightarrow\infty$ \cite{giovannetti2011,lu2020,kull2020}. According to Refs. \cite{braunstein1994,liu2020}, the Fisher information for parameter $\eta$  is maximized over all possible quantum measurement as follows,
\begin{equation}
I_\eta=4\left(\langle\partial_\eta\psi_\eta|\partial_\eta\psi_\eta\rangle-|\langle\psi_\eta|\partial_\eta\psi_\eta\rangle|^2\right),
\end{equation}
where $\psi_\eta$ represents the state carrying the information of $\eta$. The quantity $I_\eta$ is called quantum Fisher information and is independent of the specific measurements by definition.

In this paper, we explore the possibility of parameter estimation with the steady states of \nh spin chains. The paper is organized as follows. In Sec. \ref{sec_model}, we introduce the \nh spin-1/2 system and investigate the energy spectra in the presence of an external magnetic field imposed on the first site of the chain. In Sec. \ref{sec_site2}, we analytically investigate the performance of steady state parameter estimation in the \nh spin chain of two sites. In Sec. \ref{sec_siteN}, we enlarge the spin chain to finite size and numerically obtain the spectrum of the \nh system. We also discuss the dependence of quantum Fisher information on the system size. To comprehend the results, we further calculate the correlation functions of the \nh spin chain. We summarize in Sec. \ref{sec_sum}.

\section{The model and protocol}
\label{sec_model}
We consider a one-dimensional chain with nearest-neighboring pair creation interactions. On each site there is a spin-1/2 particle with upper state $|\uparrow\rangle$ and lower state $|\downarrow\rangle$. Due to the inevitable interactions with the external environment, the upper state $|\uparrow\rangle$ of each spin spontaneously decays to the lower state $|\downarrow\rangle$ in a decay rate $\gamma$.
Before the spontaneous emission from the upper state takes place, the time-evolution of the system can be governed by the effective \nh Hamiltonian as the following (set $\hbar=1$ hereinafter),
\begin{equation}
\hat{H}_{0}=J\sum_{n=1}^{N-1}{\left(\hatsigma_n^+\hatsigma_{n+1}^++\hatsigma_n^-\hatsigma_{n+1}^-\right)}-i\frac{\gamma}{4}\sum_{n}{\left(\hatsigma_n^z+\identity\right)},
\end{equation}
where $\hatsigma^\alpha_n$ ($\alpha=x,y,z$) are the spin-1/2 Pauli matrices for site $n$, $\hatsigma_n^{\pm}=\hatsigma_n^x\pm i\hatsigma_n^y$ are the creation and annihilation operators and $\identity$ is the identity operator. The parameter $J$ denotes the pair creation/annihilation strength of nearest-neighboring sites and $N$ is the number of sites in the chain.

The non-Hermitian Hamiltonian $\hat{H}_{0}$ is a specific realization of the anisotropic $XY$ Hamiltonian with the strengths of spin interactions along $x$ and $y$ directions $J_x$ and $J_y$ satisfying $J_x=-J_y$. The exact spectrum of \nh anisotropic $XY$ Hamiltonian has been discussed by Lee {\it et al} in Ref. \cite{lee2014a}. It is shown that a steady-state phase transition from the short-range order to the quasi-long-range order occurs at $J=\gamma/4$.

Now we explore the ability of the \nh chain in estimating the parameters of an external magnetic field that are imposed on the first site ($n=1$) of the chain. The external field is supposed to be restricted on the $x$-$y$ plane. The perturbed Hamiltonian is given as follows,
\begin{equation}
\hat{H}_1(h,\theta)=h\left(\cos{\theta}\hatsigma^x_1 + \sin{\theta}\hatsigma^y_1\right),
\end{equation}
where $h$ is the amplitude of the external field and $\theta$ is the azimuthal angle in the $x$-$y$ plane. We emphasize that the amplitude and the direction of the external field are time-independent. As we will see the presence of the external field will produce a nonzero stationary magnetization in the $x$-$y$ plane depending on the amplitude and direction of the external field. Our goal is to estimate the value of $h$ and $\theta$. With this goal in mind, we notice that the parameters $h$ and $\theta$ of the external field are encoded in the \nh system through modifying the energy spectrum.

Consequently, the state of the spin chain is evolved in the manner of $|\psi(t)\rangle=\exp{(-i\hat{H}t)}|\psi(0)\rangle$ where $|\psi(0)\rangle$ is the initial state. The Hamiltonian $\hat{H}$ is given by $\hat{H}=\hat{H}_0+\hat{H}_1(h,\theta)$.  Apparently, the negative imaginary parts of the eigenvalues always diminish the weight of each eigenstate in $|\psi(t)\rangle$ over time. After a sufficient amount of time, only the eigenstate associated to the eigenvalue with the largest (negative) imaginary part survives. We call such eigenstate as the steady state and label it by $|\psi_{\steadystate}\rangle$. The steady state as well as the steady-state value of the observable $\langle\hat{O}\rangle_{\steadystate}=\langle\psi_{\steadystate}|\hat{O}|\psi_{\steadystate}\rangle$ carry the information of the external field. We discuss how to estimate the parameters about the external magnetic field by measuring some observables in $|\psi_{\steadystate}\rangle$. The scheme of parameter estimation with steady states has the advantage that it does not require a precise control of the interaction time during the evolution.

\section{Two spins}
\label{sec_site2}
In order to get an intuitive picture of the parameter estimation within the \nh spin system, we first discuss the case of the two-site chain because it can be solved exactly. For $N=2$, the total Hamiltonian $\hat{H}$ are given as follows (using the open boundary condition),
\begin{equation}
\hat{H}=\left(
          \begin{array}{cccc}
            -i\gamma & 0 & he^{-i\theta} & J \\
            0 & -i\frac{\gamma}{2} & 0 & he^{-i\theta} \\
            he^{i\theta} & 0 & -i\frac{\gamma}{2} & 0 \\
            J & he^{i\theta} & 0 & 0 \\
          \end{array}
        \right).
\end{equation}
The non-Hermitian Hamiltonian admits the following equations
\begin{eqnarray}
\hat{H}|\psi_j\rangle&=&\lambda_j|\psi_j\rangle, \cr\cr
\langle\phi_j|\hat{H}&=& \lambda_j\langle\phi_j|,
\end{eqnarray}
where $|\psi_j\rangle$ and $\langle\phi_j|$ are the right and left eigenvectors that are normalized to satisfy the biorthonormality condition $\langle\phi_j|\psi_k\rangle=\delta_{jk}$, and $\lambda_j\in \mathbb{C}$ are the complex eigenvalues \cite{brody2014,wuqc2016,chenwt2022}. In particular, for the non-Hermitian systems with loss the imaginary parts of the eigenvalues are non-positive.
The eigenvalues of $\hat{H}$ can be determined by solving the secular equation $\det{(\hat{H}-\lambda I)}=0$. Thus the four eigenvalues of the total Hamiltonian $\hat{H}$ are given by
\begin{equation}
\lambda_{\pm,\pm}=-i\frac{\gamma}{2}\pm i\frac{1}{4}\left(\sqrt{\gamma^2-4J^2}\pm\sqrt{\gamma^2-4J^2-16h^2}\right).
\label{eig2}
\end{equation}

\begin{figure}[h]
  \includegraphics[width=1\linewidth]{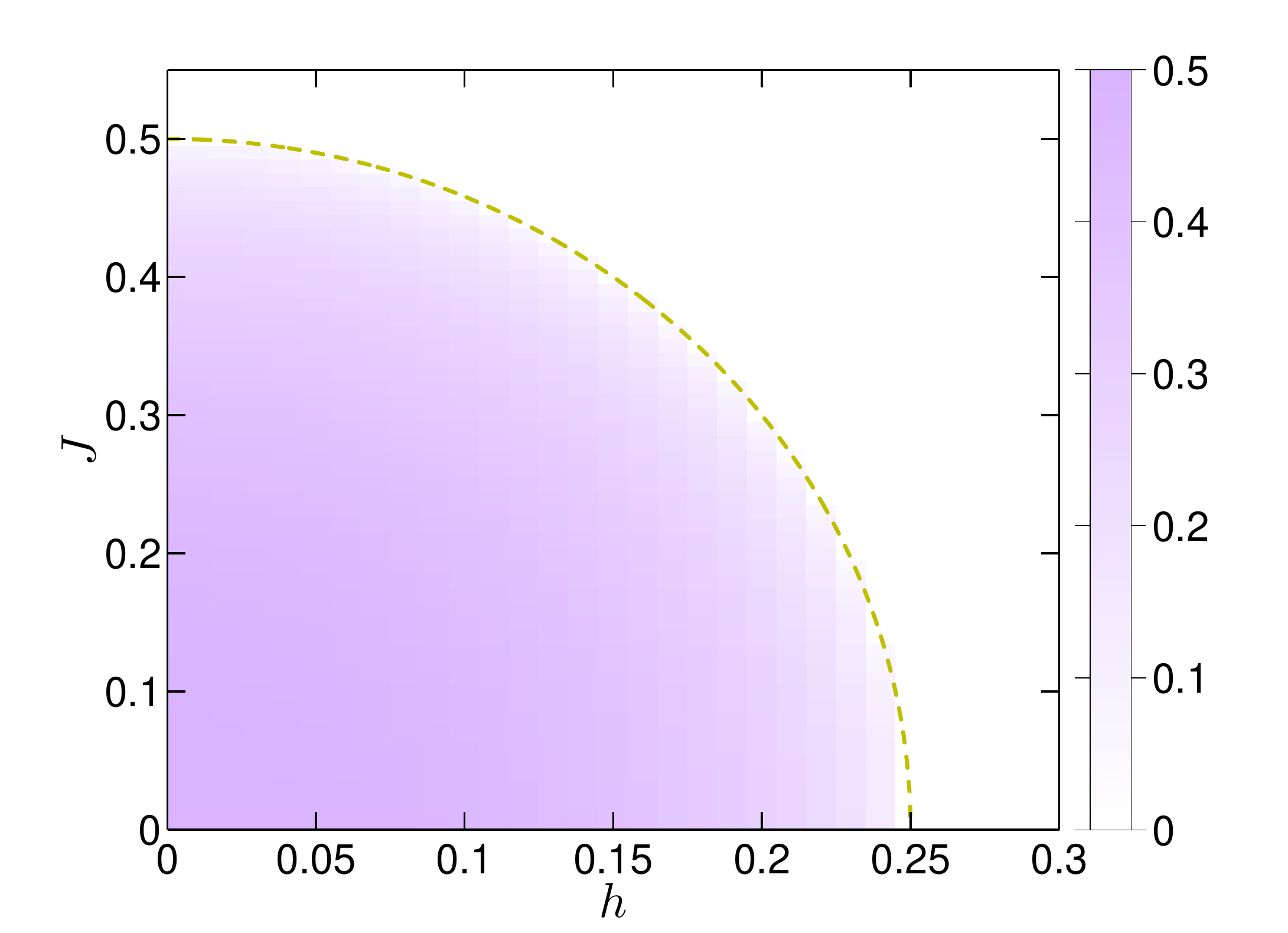}
    \caption{(Color online) The gap of imaginary parts of eigenvalues for the two-site chain. The dashed line denotes EPs of the \nh Hamiltonian, that is the gap closes at $h=\sqrt{\gamma^2-4J^2}$. The parameter is chosen as $\gamma=1$.}
     \label{Fig_gap2}
\end{figure}

From Eq. (\ref{eig2}) one can find that the imaginary parts of the eigenvalues are non-positive and irrelevant to $\theta$. We define the gap as the difference between the largest and second largest imaginary parts of the eigenvalues.
In Fig. \ref{Fig_gap2} we show the gap in the $J$-$h$ plane. It can be seen that gap closes when $h=\sqrt{\gamma^2-4J^2}$ indicating the coalescence of the eigenvalues.

In the gapped region, i.e. $h<\sqrt{\gamma^2-4J^2}/4$ or $J<\sqrt{\gamma^2-16h^2}/2$, the eigenvalue $\lambda_{+,+}$ has the largest negative imaginary part. Therefore in this region the eigenstate associated to $\lambda_{+,+}$ is the steady state,
\begin{equation}
    \vert \psi_{\steadystate}\rangle = \frac{J}{\sqrt{\gamma A}}
    \begin{pmatrix}
        e^{-i\pi/4}\sqrt{\frac{A+B}{\gamma+A}} \\
        e^{-i\left(\pi/4+\theta\right)}\sqrt{\frac{A-B}{\gamma-A}} \\
        -e^{i\left(\pi/4+\theta\right)}\sqrt{\frac{A-B}{\gamma+A}} \\
        e^{i\pi/4}\sqrt{\frac{A+B}{\gamma-A}}
    \end{pmatrix},
    \label{psi_ss}
\end{equation}
where $A=\sqrt{\gamma^2-4J^2}$ and $B=\sqrt{\gamma^2-4J^2-16h^2}$. The steady-state magnetizations of each spin can thus be calculated as the followings,
\begin{eqnarray}
\langle\hatsigma^x_1\rangle_{\steadystate}=\frac{4h}{\gamma}\cos{\left(\theta+\frac{\pi}{2}\right)},&&\langle\hatsigma^x_2\rangle_{\steadystate}=0,\cr\cr
\langle\hatsigma^y_1\rangle_{\steadystate}=\frac{4h}{\gamma}\sin{\left(\theta+\frac{\pi}{2}\right)},&&\langle\hatsigma^y_2\rangle_{\steadystate}=0,\cr\cr
\langle\hatsigma^z_1\rangle_{\steadystate}=-\frac{\sqrt{\gamma^2-4J^2-16h^2}}{\gamma},&&\langle\hatsigma^z_2\rangle_{\steadystate}=-\frac{\sqrt{\gamma^2-4J^2}}{\gamma}.
\label{mag_ss}
\end{eqnarray}
From Eq. (\ref{mag_ss}) one can see that the external magnetic field only modifies the steady-state magnetization of site $1$ and leave the steady-state magnetization of site $2$ as when the external field is absent. Although the presence of external field does not affect the steady-state magnetization of site $2$, it generates the steady-state correlations between sites $1$ and $2$ as follows,
\begin{equation}
\langle\hatsigma_1^x\hatsigma_2^x\rangle_{\steadystate}=\langle\hatsigma_1^y\hatsigma_2^y\rangle_{\steadystate}=\frac{J\sin{(2\theta)}}{\gamma}\left(1-\sqrt{\frac{B}{A}}\right),
\end{equation}
and
\begin{equation}
\langle\hatsigma_1^z\hatsigma_2^z\rangle_{\steadystate}=\frac{B}{A}.
\end{equation}

Now we focus on the steady-state magnetizations of site $1$ in which the parameter $h$ and $\theta$ are encoded in. In the absence of the external field ($h=0$), the spin is fully polarized down to the $z$-direction. According to Eq. (\ref{mag_ss}), as the external magnetic field is switched on, the magnitude of steady-state magnetization on the $x$-$y$ plane responses linearly to $h$ for a fixed $J$, and the slope reveals the information of $\theta$. The direction of the response steady-state magnetization is perpendicular to the external field. Therefore the parameters $h$ and $\theta$ can be estimated by the steady-state magnetizations.

\begin{figure}[h]
  \includegraphics[width=1\linewidth]{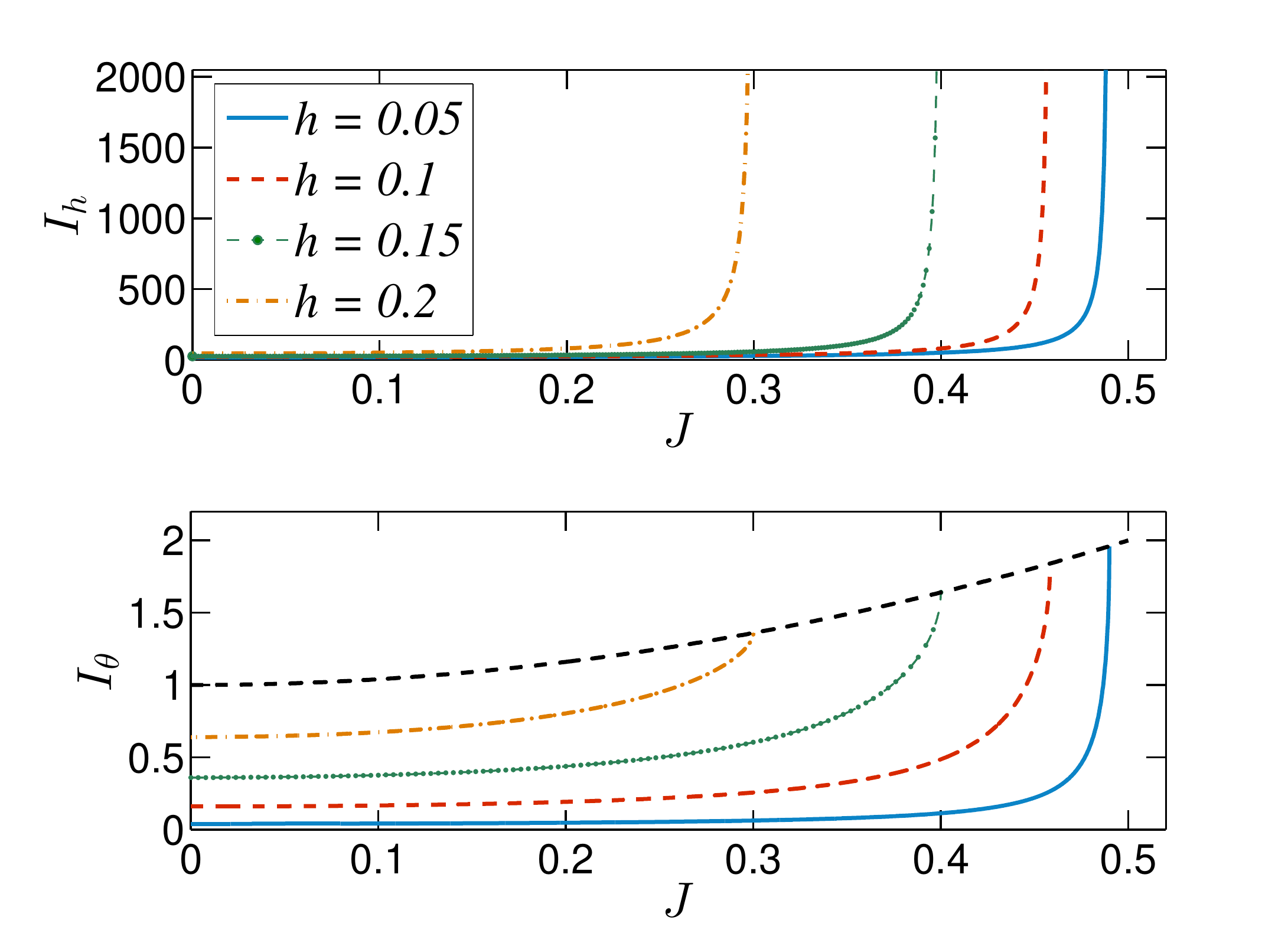}
   \caption{(Color online) The quantum Fisher information $I_h$ (top panel) and $I_\theta$ (bottom panel) a functions of $J$ for different values of $h$. The black dashed line is the bound of the maximal quantum Fisher information about $\theta$ for fixed $h$. The parameter is chosen as $\gamma=1$.}
  \label{Fig_qfi2}
\end{figure}

By virtue of Eq. (\ref{psi_ss}), the quantum Fisher information about $h$ and $\theta$ associated to the steady states are given by
\begin{equation}
I_h=\frac{16}{\gamma^2-4J^2-16h^2},
\end{equation}
and
\begin{equation}
I_\theta=\frac{(A-B)(AB+\gamma^2+4J^2)}{\gamma A}.
\end{equation}
In Fig. \ref{Fig_qfi2}, we show the quantum Fisher information as functions of $J$ for different values of $h$. For fixed $h$, one can see that both $I_h$ and $I_\theta$ increase as the coupling strength approaching to the EP $J_c=\sqrt{\gamma^2-16h^2}/2$. In particular, at the EP the quantum Fisher information $I_h$ diverges while the $I_\theta$ reaches to the maximum $I^{\rm max}_\theta=1+4J^2/\gamma^2$. The results reveal that the EP of the \nh spin chain indeed enhances the performance of parameter estimation with the steady states.

\section{Finite-size chain}
\label{sec_siteN}
\begin{figure}[h]
  \includegraphics[width=1\linewidth]{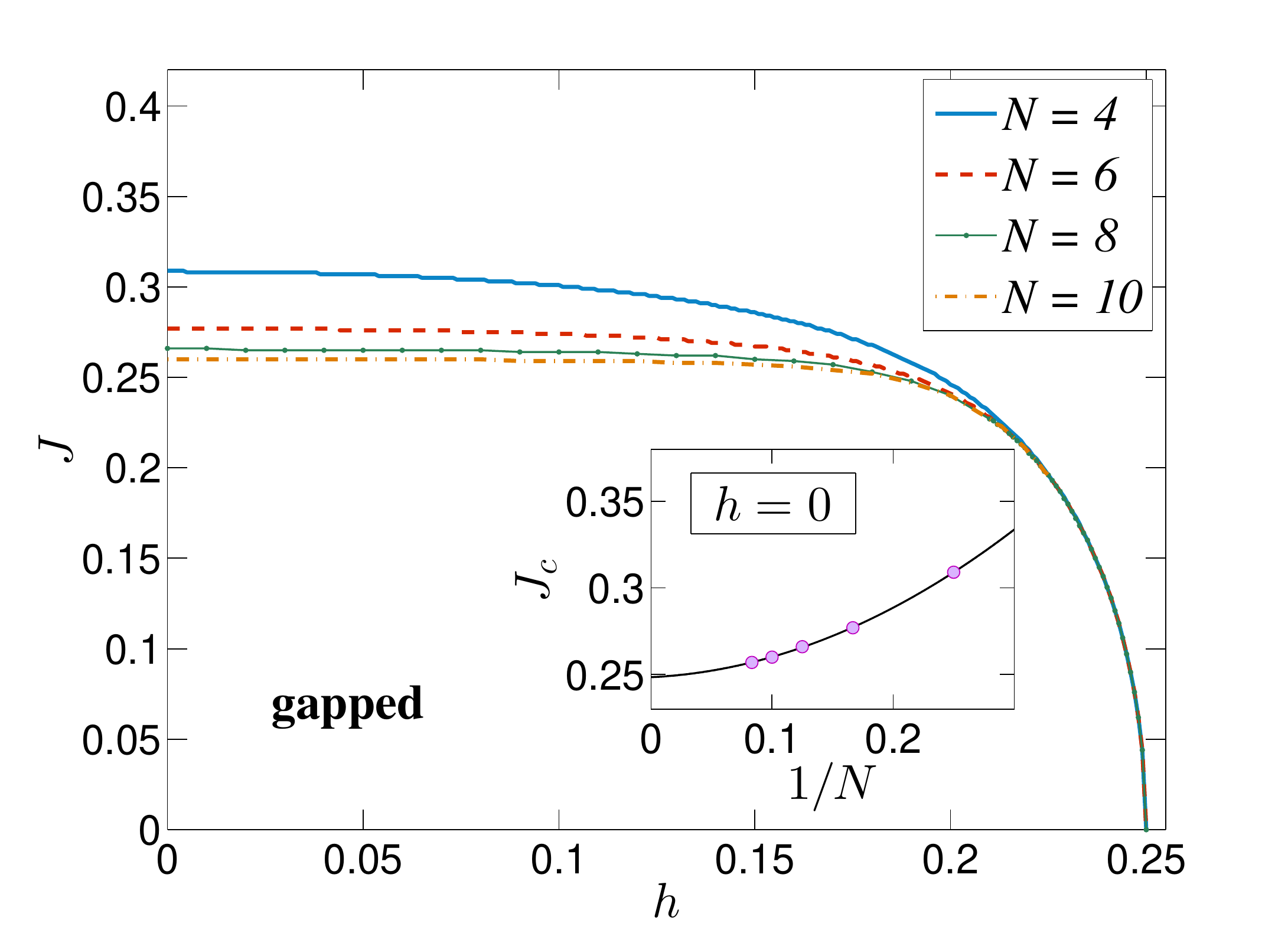}
    \caption{(Color online) The EPs of \nh Hamiltonian for finite-size chains with different $N$. For the parameters in the left-bottom region of the main panel the imaginary part of the eigenvalues in the gapped. In the inset, it is shown the fitting of EP to the inverse of system size for $h=0$. The solid (black) line is the fitting result $J_c(N)=0.842/N^2+0.031/N+0.249$. The parameter is chosen as $\gamma=1$.}
     \label{Fig_gapN}
\end{figure}

Now we enlarge the size of the spin chain to the cases of $N>2$. The presence of external field in the \nh Hamiltonian prevents us to obtain the analytical energy spectra for finite-size chains, we leave it to the numerical diagonalization of $\hat{H}$ (using open boundary condition). In Fig. \ref{Fig_gapN}, we show the EPs for the \nh Hamiltonians for spin chains with different sizes $N$. The Hamiltonians with the parameters in the regions on the left-bottom of the curves are the gapped. Compared to the case of $N=2$, the gapped region shrinks quantitatively as the size of spin chain increasing and tends to converge for a sufficiently long chain. The trends of convergence have already been observed up to $N=10$ for $h\gtrsim 0.18\gamma$. For small $h$, the relatively long chains are needed to reveal the convergency. In the limit case of $h=0$, we fit the EPs $J_c$ to the inverse of $N$ in the inset of Fig.  \ref{Fig_gapN}. The EP for infinite-long ($1/N\rightarrow 0$) spin chain is extrapolated as $J_c\approx 0.249\gamma$ which is in good agreement with the exact result $J_c=\gamma/4$ by Jordan-Wigner transformation \cite{lee2014a}.

\begin{figure}[h]
  \includegraphics[width=1\linewidth]{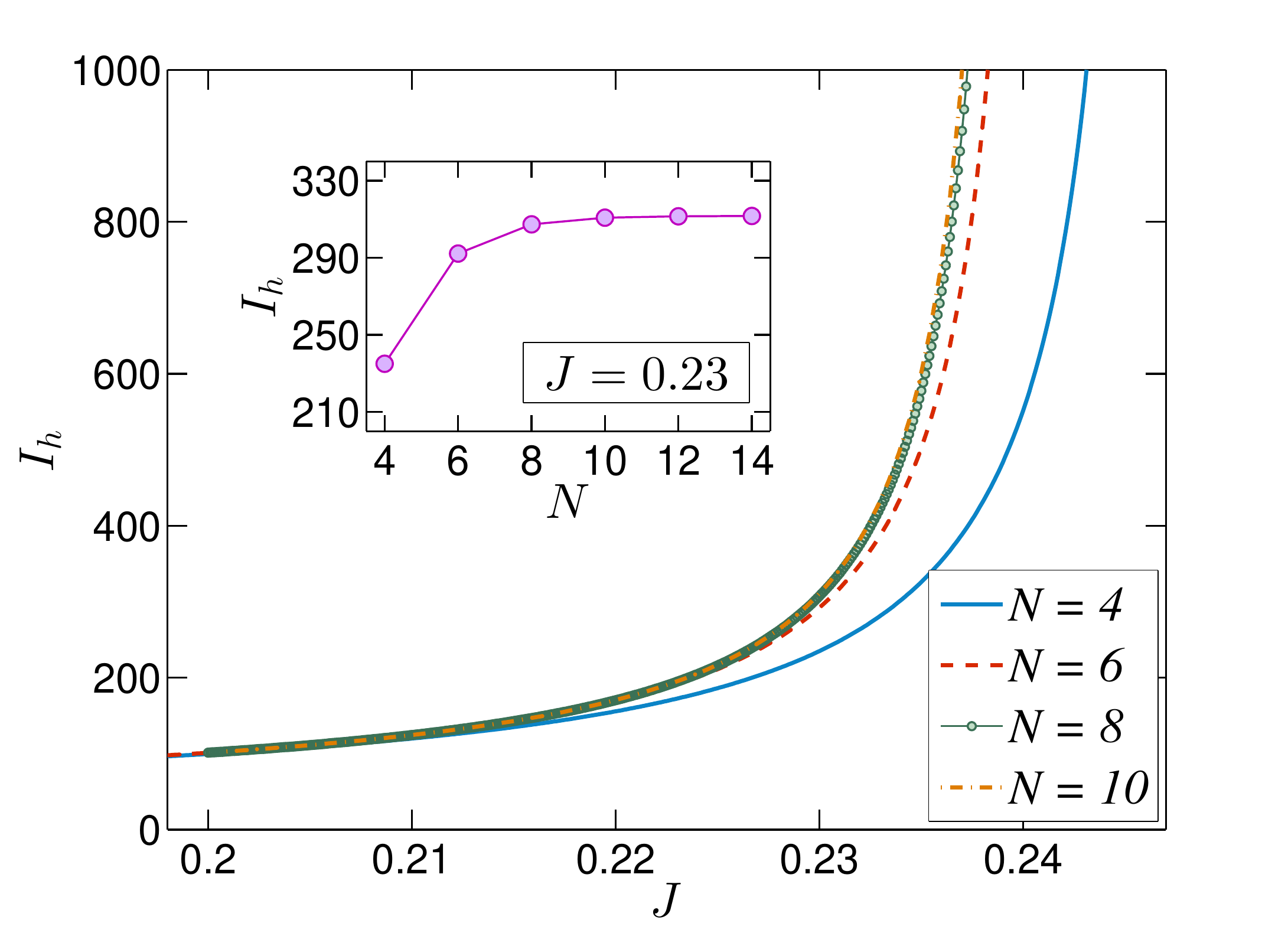}
    \caption{(Color online) The quantum Fisher information about the amplitude of the external magnetic field for spin chains of different sizes.  The parameter are chosen as $h=0.2$, $\theta=0$ and $\gamma=1$.}
     \label{Fig_qfiN}
\end{figure}

The quantum Fisher information about the amplitude of the external field for different sizes of spin chains is shown in Fig. \ref{Fig_qfiN}. Similar to the case of $N=2$, one can find that the quantum information increase dramatically as the coupling strength approaches to the EPs for each $N$. For fixed $J$, it is interesting that the quantum Fisher information increases as the system size enlarges and saturates for sufficiently long chain. For instance, in the inset of Fig. \ref{Fig_qfiN} it is shown the dependence of $I_h$ on system size $N$. The quantum Fisher information saturates up to $N=14$ for $J=0.23$ and $h=0.2$.

\begin{figure}[h]
  \includegraphics[width=1\linewidth]{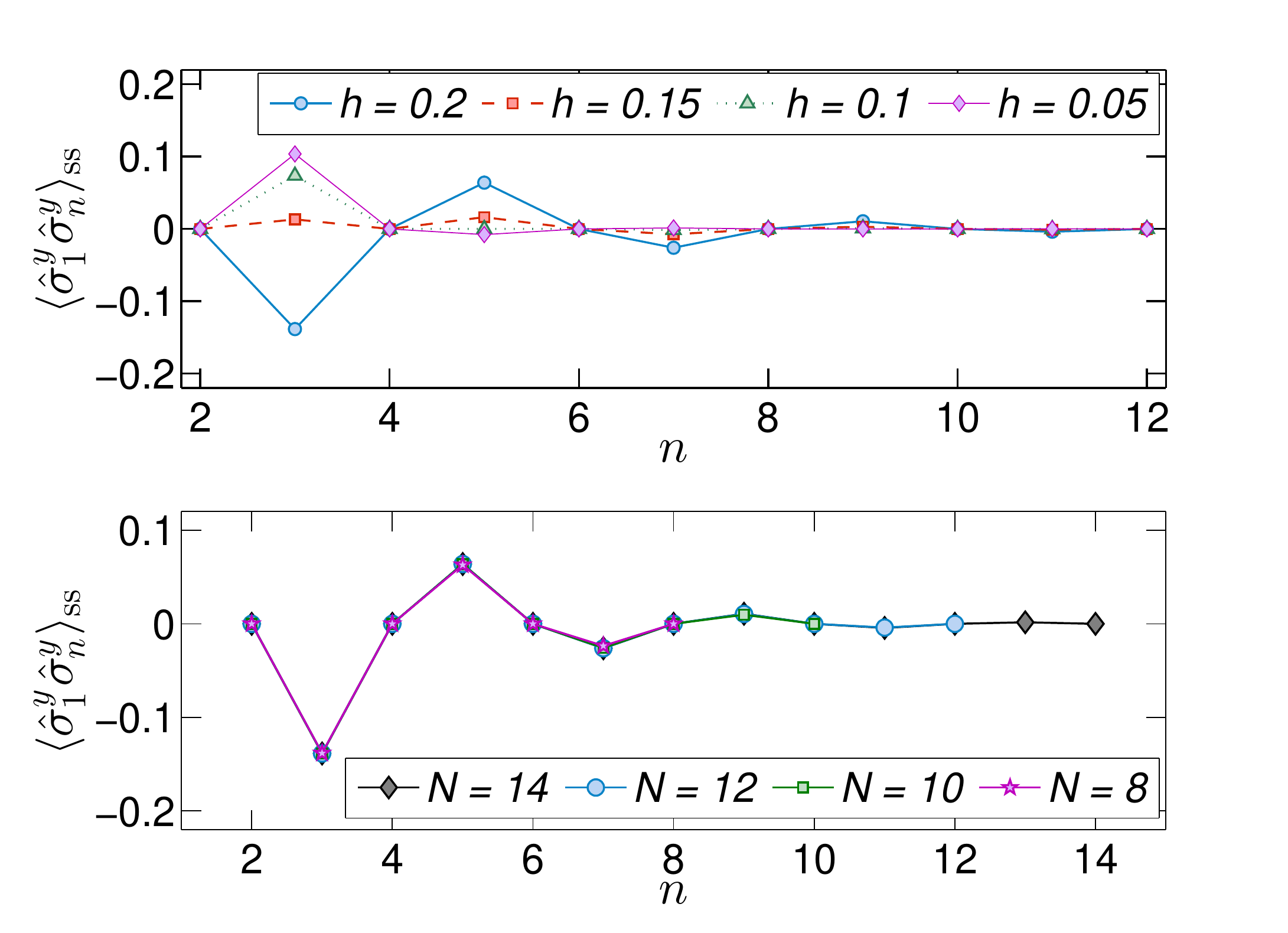}
    \caption{(Color online) Top panel: The steady-state correlations $\langle\hatsigma_1^y\hatsigma_{n}^y\rangle_{\steadystate}$ for different $h$ in a spin chain of size $N=12$.  Bottom panel:  Steady-state correlations in spin chains of different sizes for $h=0.2$. Other parameters are chosen as $J=0.23$, $\theta=0$ and $\gamma=1$.}
     \label{Fig_corr}
\end{figure}

The saturation of $I_h$ as system size increases is in contrast to the quantum sensing protocol with steady-state phase transitions of dissipative systems \cite{raghunandan2018}.
In the latter protocol, the external field is imposed on all the spins of the system. In thermodynamic limit ($N\rightarrow\infty$) the steady-state phase transition occurs at the critical amplitude of imposing field while a strong correlation is created through the whole lattice. Here in our protocol, the external field is imposed on the first site of the chain. We show the steady-state correlations $\langle\hatsigma_1^y\hatsigma_{n}^y\rangle_{\steadystate}$ ($n\ge 2$) for a spin chain of size $N=12$ in the top panel of Fig. \ref{Fig_corr}. One can see that for $J=0.23$ the correlations become stronger when $h$ gets close to the EP. However, the correlations decay to zero within a short range, as shown in the bottom panel of Fig. \ref{Fig_corr} where the vanishing of $\langle\hatsigma_1^y\hatsigma_{n}^y\rangle_{\steadystate}$ is confirmed by the chains up to $N=14$.

\section{Summary}
\label{sec_sum}
In summary, we have discussed the possibility of parameter estimation in a one-dimensional \nh spin1-/2 chain with pair creations/annihilations. The parameters to be estimated are the amplitude and the azimuthal angle of an external magnetic field that imposed on the first site of the chain. We presented the analytical spectrum for the chain with two sites as well as the steady-state magnetizations for each site. We found that the quantum Fisher information about the amplitude diverges near the EPs while the quantum Fisher information about $\theta$ reaches to the maximum at EPs. For the finite-size chain ($N>2$), we numerically obtained the gapped regions of the imaginary parts of eigenvalues by finite-size scaling. Similar to the case of $N=2$, the quantum Fisher information $I_h$ diverges at EPs. Moreover, the quantum Fisher information converges to a finite value as the length of the chain increases because only short range correlations are created in the presence of the external magnetic field.

Our work shows that the \nh quantum many-body system may serve as a parameter estimator. In particular, the performance will be enhanced near the EPs of the eigenvalues. However, for the specific model considered here, because the presence of the external field modifies the spectrum of the \nh spin chain and even eliminates the EP for large $h$, the range of parameters that can be efficiently estimated shrinks into a finite region ($h<0.25$).

Our discussion is restricted in the regions that the imaginary parts of the spectrum are gapped. As the coupling strength crosses the EP, the eigenstates and the eigenvalues of Hamiltonian coalesce and the imaginary part of the spectrum becomes gapless. For future work, one may develop new tools for quantifying the quantum Fisher information basing on the coalesced steady states. It is also interesting to investigate the response of magnetizations of the non-Hermitian spin chains in the gapless regions of the spectrum.

We finish with a brief discussion on the experimental implementation of our scheme. The many-body Hamiltonian can be realized in an array of coupled cavities \cite{fitzpatrick2017,collodo2019} and ensembles of atoms \cite{weimer2010,bluvstein2021}. These platforms have the nice flexibility to tuning the interactions as well as the coupling strengths. For example, the coupling strength can be tuned by controlling the intensities of the lasers that imposed on the atoms \cite{hartmann2007} and the non-Hermitian parts of the Hamiltonian is implementable with the help of the auxiliary level \cite{lee2014a}.

\acknowledgments
This work is supported by National Natural Science Foundation of China under Grant No. 11975064.

\end{document}